\RequirePackage{lineno}
\documentclass[aps,twocolumn,showpacs,byrevtex,prl,reprint]{revtex4-1}

\usepackage{graphicx}
\usepackage{dcolumn}
\usepackage{bm}
\usepackage{rotating}
\usepackage{color}
\usepackage{verbatim} 
\usepackage{multirow}

\newcommand{\PreserveBackslash}[1]{\let\temp=\\#1\let\\=\temp}
\newcolumntype{C}[1]{>{\PreserveBackslash\centering}p{#1}}
\newcolumntype{R}[1]{>{\PreserveBackslash\raggedleft}p{#1}}
\newcolumntype{L}[1]{>{\PreserveBackslash\raggedright}p{#1}}

\newcommand{\EE}{e^+e^-}

\newcommand{\st}{\sqrt{s}=4.23~\rm{GeV}}
\newcommand{\sd}{\sqrt{s}=4.26~\rm{GeV}}

\begin{document}


\title{\boldmath Study of $e^+e^-\to\omega\chi_{cJ}$ at center-of-mass energies from 4.21 to 4.42~GeV}

\author{
  \begin{small}
    \begin{center}
      M.~Ablikim$^{1}$, M.~N.~Achasov$^{8,a}$, X.~C.~Ai$^{1}$,
      O.~Albayrak$^{4}$, M.~Albrecht$^{3}$, D.~J.~Ambrose$^{42}$,
      A.~Amoroso$^{46A,46C}$, F.~F.~An$^{1}$, Q.~An$^{43}$, J.~Z.~Bai$^{1}$
% ...
      \\
      \vspace{0.2cm}
      (BESIII Collaboration)\\
      \vspace{0.2cm} 
      { \it
        $^{1}$ Institute of High Energy Physics, Beijing 100049, People's Republic of China\\
% ...
$^{3}$ Bochum Ruhr-University, D-44780 Bochum, Germany\\
        $^{4}$ Carnegie Mellon University, Pittsburgh, Pennsylvania 15213, USA\\
% ...
$^{8}$ G.I. Budker Institute of Nuclear Physics SB RAS (BINP), Novosibirsk 630090, Russia\\
% ...
$^{42}$ University of Rochester, Rochester, New York 14627, USA\\
        $^{43}$ University of Science and Technology of China, Hefei 230026, People's Republic of China\\
% ...
$^{46}$ (A)University of Turin, I-10125, Turin, Italy; (B)University of Eastern Piedmont, I-15121, Alessandria, Italy; (C)INFN, I-10125, Turin, Italy\\
% ...
$^{a}$ Also at the Novosibirsk State University, Novosibirsk, 630090, Russia\\
% ...
      }\end{center}
    \vspace{0.4cm}
  \end{small}
}

\affiliation{}

\begin{abstract}

Based on data samples collected with the BESIII detector at the
BEPCII collider at 9 center-of-mass energies from 4.21 to
4.42~GeV, we search for the production of $e^+e^-\to
\omega\chi_{cJ}$ ($J=0$, 1, 2). The process $e^+e^-\to
\omega\chi_{c0}$ is observed for the first time, and the Born
cross sections at $\sqrt{s}=4.23$ and 4.26~GeV are measured to be
$(55.4\pm 6.0\pm 5.9)$ and $(23.7\pm 5.3\pm 3.5)$~pb,
respectively, where the first uncertainties are statistical and
the second are systematic.
The $\omega\chi_{c0}$ signals
at the other 7 energies and $e^+e^-\to
\omega\chi_{c1}$ and $\omega\chi_{c2}$ signals are not
significant, and the upper limits on the
cross sections are determined.
By examining the $\omega\chi_{c0}$ cross section
as a function of center-of-mass energy,
we find that it is inconsistent with the line shape of the $Y(4260)$
observed in $e^+ e^-\to\pi^+\pi^-J/\psi$.
Assuming the $\omega\chi_{c0}$ signals
come from a single resonance, we extract
mass and width of the resonance to
be $(4230\pm8\pm6)$~MeV/$c^2$ and $(38\pm12\pm2)$ MeV, respectively,
and the
statistical significance is more than $9\sigma$.
\end{abstract}


\maketitle

The charmonium-like state $Y(4260)$ was first observed in its decay to
$\pi^+\pi^- J/\psi$~\cite{babar-y4260-1}, and its decays into $\pi^0\pi^0 J/\psi$ and $K^+K^- J/\psi$
were reported from a study of 12.6~${\rm pb}^{-1}$ data collected at 4.26 GeV by the CLEO-c
experiment~\cite{cleoc}.
Contrary to the hidden charm final states, the $Y(4260)$ were found to have small coupling to
open charm decay modes~\cite{belle-dd-1}, as well as to light hadron final states~\cite{light1,light2}.
Recently, charged charmoniumlike states
$Z_c(3900)$ [$\pi^{\pm}J/\psi$]~\cite{Zc-bes,Zc-belle,Zc-cleo},
$Z_c(3885)$ [$(D\bar{D}^*)^\pm$]~\cite{Zc-xuxp}, $Z_c(4020)$ [$(\pi h_c)$]~\cite{Zc-4020,Zc-jiqp},
and
$Z_c(4025)$ [$(D^*\bar{D}^*)^\pm$]~\cite{Zc-lvxr} were observed in $e^+e^-$ data collected
around $\sqrt{s}=4.26$ GeV.
These features suggest the existence of a complicated
substructure of the $Y(4260)\to\pi^+\pi^-J/\psi$ as well as the nature of the $Y(4260)$ itself.
Searches for new decay modes
and measuring the line shape may provide information that is
useful for understanding the nature of the $Y(4260)$.

Many theoretical models have been proposed to interpret the $Y(4260)$,
{\it e.g.}, as a quark-gluon charmonium hybrid, a tetraquark
state, a
hadro-charmonium,
or a hadronic molecule~\cite{QWG}.
The authors of Ref.~\cite{omg-chi0} predict a sizeable coupling
between the $Y(4260)$ and the $\omega\chi_{c0}$ channel by
considering the threshold effect of $\omega\chi_{c0}$ that plays
a role in reducing the decay rates into open-charm channels.
By adopting the spin rearrangement scheme in the heavy
quark limit and the experimental information, Ref.~\cite{Ratio}
predicts the ratio of the decays $Y(4260)\to\omega\chi_{cJ}
~(J=0,1,2)$ to be $4:3:5$.

In this Letter, we report on the study of $e^+e^-\to
\omega\chi_{cJ}~(J=0,~1,~2)$ based on the $e^+e^-$ annihilation
data samples collected with the BESIII detector~\cite{S-ring}
at 9 center-of-mass energy
points in the range $\sqrt{s}=4.21-4.42$~GeV.
In the analysis,
the $\omega$ meson is reconstructed via its $\pi^+\pi^-\pi^0$ decay mode,
the $\chi_{c0}$ state is via $\pi^+\pi^-$ and $K^+K^-$ decays,
and the $\chi_{c1,2}$ states are via
$\chi_{c1,2}\to\gamma J/\psi$,
$J/\psi\to\ell^+\ell^-~(\ell=e,~\mu)$.

We select charged tracks, photon, and $\pi^0\to\gamma\gamma$
candidates as described in Ref.~\cite{bianjm}.
A candidate event must have four tracks with zero net charge and
at least one $\pi^0$ candidate; for the $e^+e^-\to \omega\chi_{c1,2}$
channels, an additional photon is required. The tracks with a
momentum larger than $1~{\rm GeV}/c$ are identified as originating
from $\chi_{cJ}$, lower momentum pions are interpreted as originating
from $\omega$ decays. A 5C
kinematic fit is performed to constrain the total four-momentum of
all particles in the final states to that of the initial $e^+e^-$
system, and $M_{\gamma\gamma}$ is
constrained to $m_{\pi^0}$. If more than one
candidate occurs in an event, the one with the smallest
$\chi^2_{\rm 5C}$ of the kinematic fit is selected. For the channel
$e^+e^-\to \omega\chi_{c0}$, the two tracks from the $\chi_{c0}$ are
assumed to be $\pi^+\pi^-$ or $K^+K^-$ pairs. If
$\chi^2_{\rm 5C}(\pi^+\pi^-) < \chi^2_{\rm 5C}(K^+K^-)$, the event is
identified as originating from the $\pi^+\pi^-$ mode, otherwise it
is considered to be from the $K^+K^-$ mode. $\chi^2_{\rm 5C}$ is required to be
less than 100.
For the $J/\psi$ reconstruction, 
the charged particle with the energy deposition in ECL larger than 1~GeV is identified
as $e$, otherwise it is $\mu$.
The $\chi^2_{\rm 5C}$ for the $\omega\chi_{c1,2}$ candidate event is
required to be less than 60.

\begin{figure}[htbp]
\begin{center}
\includegraphics[width=0.48\textwidth]{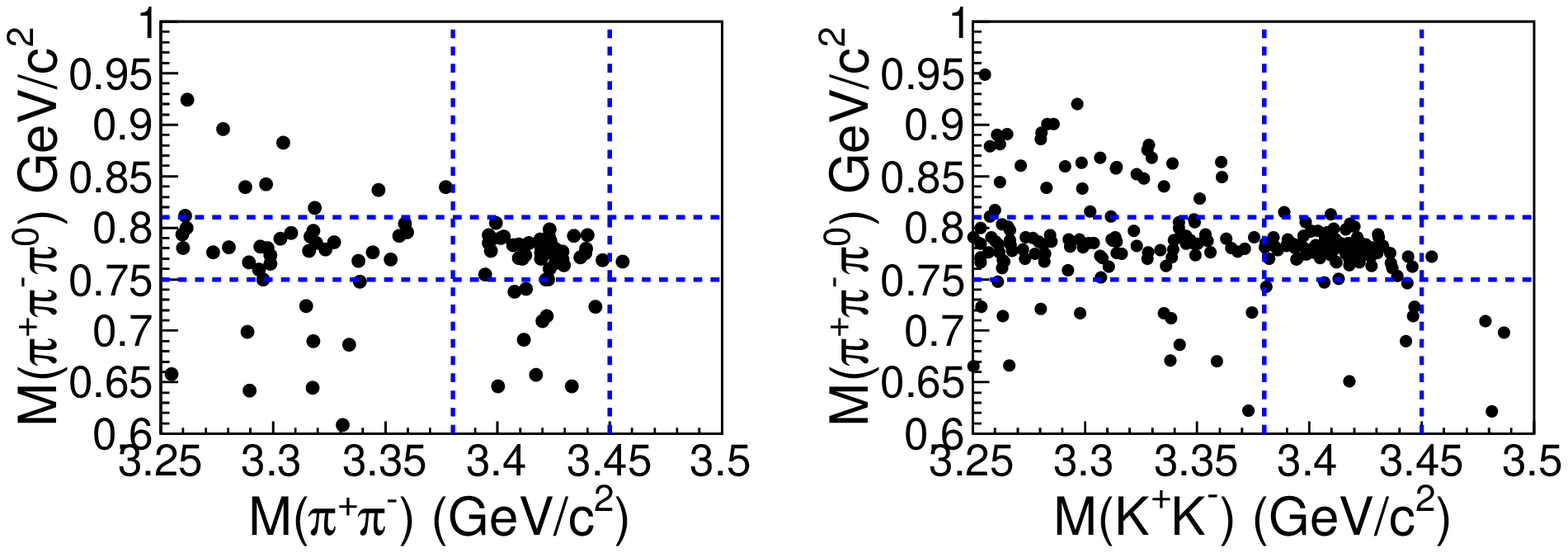}
\includegraphics[width=0.48\textwidth]{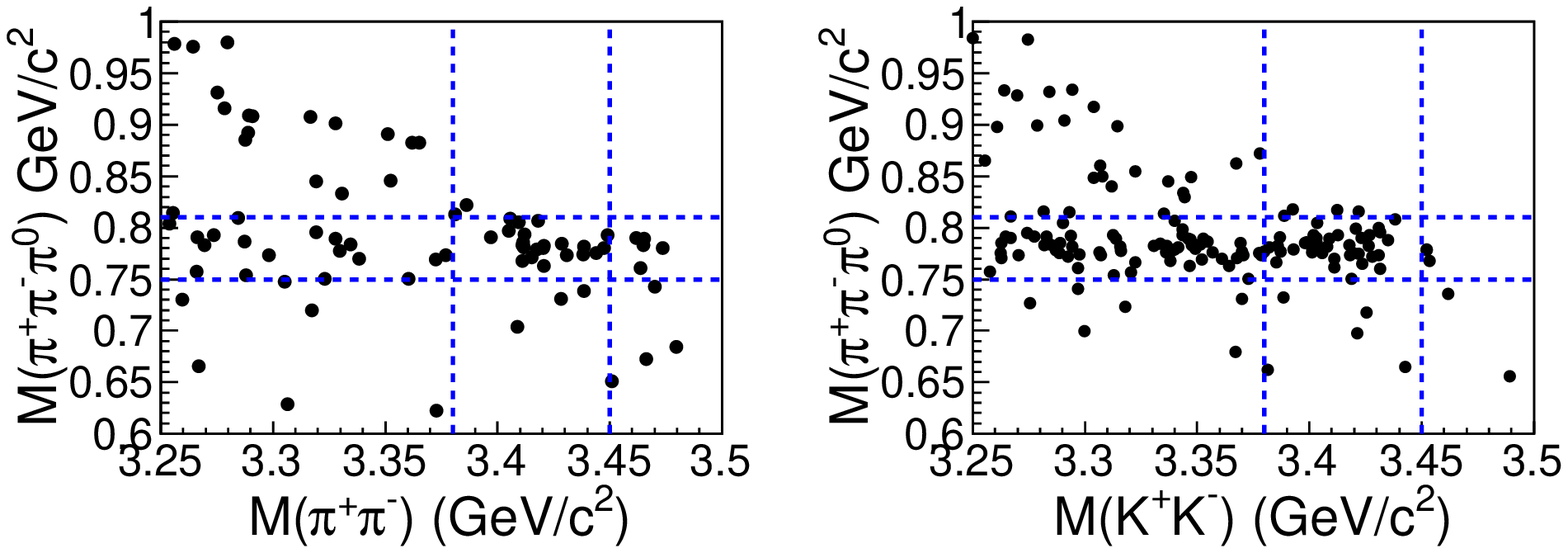}
\caption{Scatter plots of the $\pi^+\pi^-\pi^0$ invariant mass
versus the $\pi^+\pi^-$ (left) and $K^+K^-$ (right) invariant mass at
$\st$ (top) and 4.26 GeV (bottom). The dashed lines denote the
$\omega$ and $\chi_{c0}$ signal regions. }
\label{2D-omg-chic0}
\end{center}
\end{figure}

The main sources of background after event selection are
found to be $e^+e^-\to \omega\pi^+\pi^-(\omega K^+K^-)$, where
the $\pi^+\pi^-(K^+K^-)$ are not from $\chi_{c0}$ decays. The scatter
plots of the invariant mass of $\pi^+\pi^-\pi^0$
versus that of
$\pi^+\pi^-$ or $K^+K^-$ for data at $\sqrt{s}=4.23$ and 4.26~GeV
are shown in Fig.~\ref{2D-omg-chic0}. Clear accumulations of
events are seen around the intersections of the $\omega$ and
$\chi_{c0}$ regions, which indicate $\omega\chi_{c0}$ signals.
Signal candidates are
required to be in the $\omega$ signal region [0.75, 0.81] GeV/$c^2$,
The $\omega$ sideband is taken as [0.60, 0.72] GeV/$c^2$
to estimate the non-resonant background.

\begin{figure}[htbp]
\begin{center}
\includegraphics[width=0.48\textwidth]{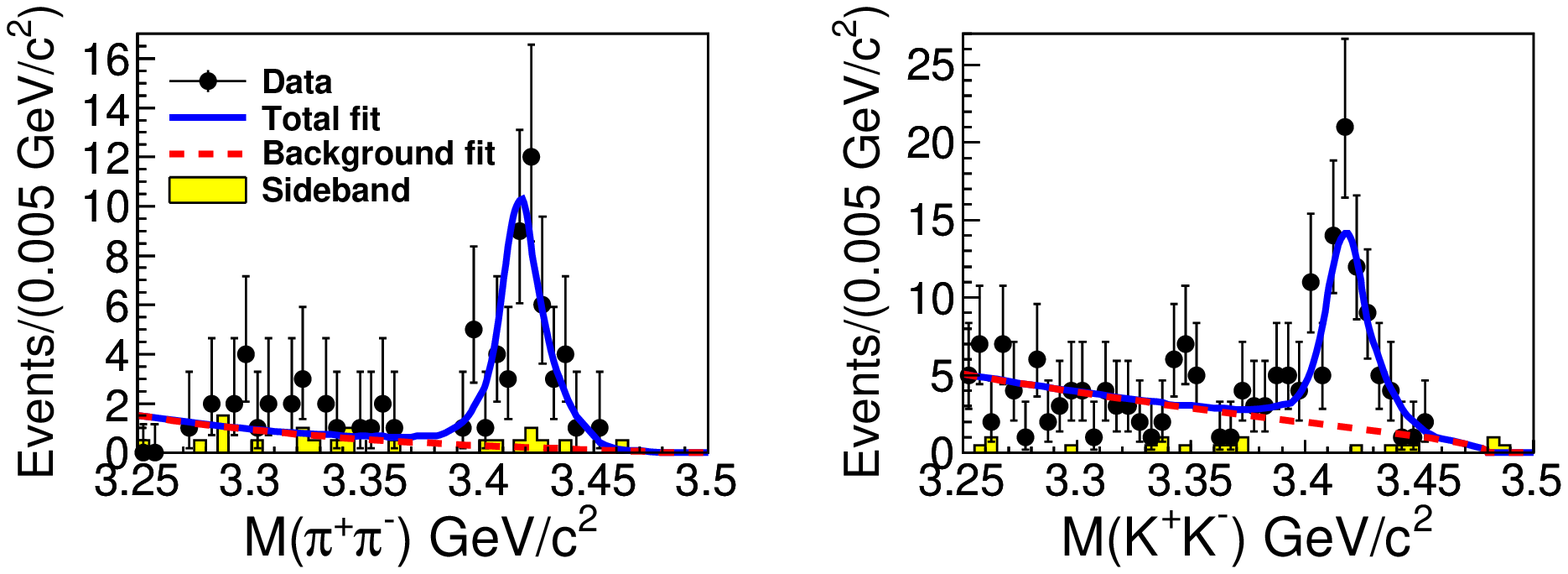}
\includegraphics[width=0.48\textwidth]{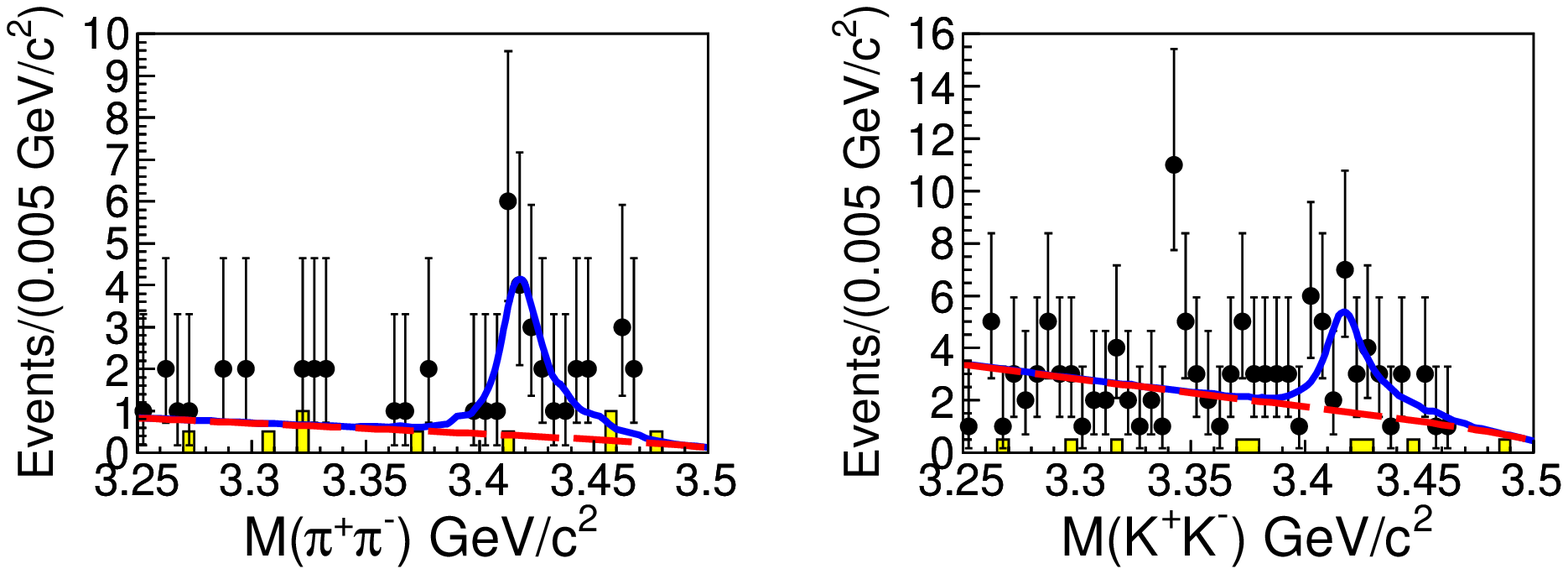}
\caption{Fit to the invariant mass distributions $M(\pi^+\pi^-)$
(left) and $M(K^+K^-)$ (right) after requiring
$M(\pi^+\pi^-\pi^0)$ in the $\omega$ signal region
at $\st$ (top) and 4.26 GeV (bottom). Points with error
bars are data,
the solid curves are the fit results, the dashed lines indicate
the background and the shaded histograms
show the normalized $\omega$ sideband events.
} \label{Fit4230}
\end{center}
\end{figure}

Figure~\ref{Fit4230} shows $M(\pi^+\pi^-)$ and $M(K^+K^-)$ at
$\sqrt{s}=4.23$ and 4.26~GeV after all requirements are
imposed. To extract the signal yield, an
unbinned maximum likelihood fit is performed on the $\pi^+\pi^-$
and $K^+K^-$ modes simultaneously. The signal is described with a
shape determined from the simulated signal MC sample. The
background is described with an ARGUS function,
$m\sqrt{1-(m/m_0)^2}\cdot e^{k(1-(m/m_0)^2)}$~\cite{argus},
where $k$ is a free parameter in the
fit, and $m_0$ is fixed at $\sqrt{s}-0.75$~GeV (0.75 GeV is the
lower limit of the $M(\pi^+\pi^-\pi^0)$ requirement).
In the fit,
the ratio of the number of $\pi^+\pi^-$ signal events to that of $K^+K^-$
signal events is fixed to be $\frac{\epsilon_{\pi} \mathcal{B}(\chi_{c0}\to
\pi^+\pi^-)} {\epsilon_K\mathcal{B}(\chi_{c0}\to K^+K^-)}$, where
$\mathcal{B}(\chi_{c0}\to \pi^+\pi^-)$ and
$\mathcal{B}(\chi_{c0}\to K^+K^-)$ are taken as world average
values~\cite{PDG}, and $\epsilon_\pi$ and $\epsilon_K$ are the
efficiencies of $\pi^+\pi^-$ and $K^+K^-$ modes determined from MC
simulations, respectively. The possible interference between the
signal and background is neglected. The fit results
are shown in Fig.~\ref{Fit4230}.
For the $\sqrt{s}=4.23$~GeV data, the total signal
yield of the two modes is $125.3\pm 13.5$, and the signal statistical
significance is $11.9\sigma$. By projecting the events
of the two modes into
two histograms (at least 7 events per bin), the goodness-of-fit is
found to be
$\chi^2/\rm {d.o.f.} = 37.6/22$, where the {\rm d.o.f.}~is the number of degrees of
freedom. For the $\sd$ data, the total signal yield is $45.5\pm
10.2$ with a statistical significance of $5.5\sigma$, and 
$\chi^2/\rm {d.o.f.}=27.1/15$. Since the statistics at
the other energy points are very limited, the number of the observed
events is obtained by counting the entries in the $\chi_{c0}$
signal region [3.38,~3.45]~GeV/$c^2$, and the number of
background events in the signal region is obtained by
fitting the $M(\pi^+\pi^-)$
$[M(K^+K^-)]$ spectrum excluding the $\chi_{c0}$ signal region
and scaling to the size of the signal region.

\begin{table*}[htbp]
\small
\begin{center}
\caption{The results on $e^+e^-\to \omega \chi_{c0}$. Shown in the
table are the integrated luminosity $\mathcal{L}$, product of radiative
correction factor, branching fraction and efficiency
$\mathcal{D}=(1+\delta^r)\cdot(\epsilon_{\pi}\cdot\mathcal{B}
(\chi_{c0}\to\pi^+\pi^-)+\epsilon_{K}\cdot\mathcal{B}(\chi_{c0}\to K^+K^-))$,
number of observed events $N^{\rm
{obs}}$ (the numbers of background are subtracted at $\sqrt{s}=4.23$
and 4.26 GeV), number of estimated background $N^{\rm bkg}$, vacuum polarization factor
$(1+\delta^v)$, Born cross section $\sigma^{\rm B}$, and upper
limit (at the 90\% C.L.) on Born cross section $\sigma_{\rm UL}^{\rm
B}$ at each energy point. The first uncertainty of the Born cross
section is statistical, and the second systematic. The dashes mean
not available. } \label{CrossSection-up}
\begin{tabular}{cccccrcc}
\hline
 $\sqrt{s}$ (GeV) &$\mathcal{L}~(\rm pb^{-1})$& $\mathcal{D}~(\times10^{-3})$  &  $N^{\rm {obs}}$ &  $N^{\rm bkg}$    &$1+\delta^v$& $\sigma^{\rm B}$ (pb) & $\sigma_{\rm UL}^{\rm B}$ (pb)\\
\hline
4.21    &54.6   & 1.99              & 7         &  $5.0\pm2.8$              &1.057   &$20.2^{+46.3}_{-37.7}\pm3.3$     &  $<90$\\
4.22    &54.1   & 2.12               & 7         &  $4.3\pm2.1$             &1.057   &$25.1^{+39.4}_{-30.4}\pm2.0$     &  $<81$\\
4.23    &1047.3 & 2.29         & $125.3\pm13.5$ &-                    &1.056  & $55.4\pm6.0\pm5.9$  &-      \\
4.245   &55.6   & 2.44               & 6         &  $4.0\pm1.5$              &1.056   &$16.3^{+30.8}_{-22.3}\pm1.5$     &  $<60$\\
4.26    &826.7  & 2.50        & $45.5\pm10.2$  &-                    &1.054  & $23.7\pm5.3\pm3.5$   &-  \\
4.31    &44.9   & 2.56               & 5         &  $2.2\pm1.6$             &1.053   &$26.2^{+34.9}_{-25.1}\pm2.2$     &  $<76$\\
4.36    &539.8  &2.62                 & 29       &  $32.4\pm4.7$            &1.051   &$-2.6^{+6.1}_{-5.4}\pm0.27$      &  $<6$\\
4.39    &55.2   & 2.57                 & 2         &  $0.6\pm0.7$             &1.051   &$10.4^{+20.7}_{-11.2}\pm0.7$     &  $<37$\\
4.42    &44.7   &2.46                 & 0         &  $1.4\pm1.5$             &1.053   &$-13.6^{+18.5}_{-14.7}\pm1.3$    &  $<15$\\
\hline
\end{tabular}
\end{center}
\end{table*}

For the process $e^+e^-\to \omega\chi_{c1,2}$, the main
remaining backgrounds stem from
$e^+e^-\to\pi^+\pi^-\psi',~\psi'\to\pi^0\pi^0 J/\psi$ and
$e^+e^-\to\pi^0\pi^0\psi',~\psi'\to\pi^+\pi^- J/\psi$. To suppress
these backgrounds,
we exclude events in which the invariant mass
$M(\pi^+\pi^-\ell^+\ell^-)$ or the mass recoiling against $\pi^+\pi^-$
[$M^{\rm recoil}(\pi^+\pi^-)$] lie in the region
[3.68,~3.70]~GeV/$c^2$.

The $J/\psi$ and $\omega$ signal regions are set to be
[3.08,~3.12]~GeV/$c^2$ and [0.75,~0.81]~GeV/$c^2$, respectively.
After all the requirements are applied, no obvious signals are
observed at $\sqrt{s}=4.31,~4.36,~4.39,$ and 4.42~GeV. The number
of observed events is obtained by counting events in the
$\chi_{c1}$ or $\chi_{c2}$ signal regions, which are defined as
[3.49,~3.53] or [3.54,~3.58]~GeV/$c^2$, respectively. The number
of background events in the signal regions is estimated with
data obtained from the sideband
region [3.35,~3.47]~GeV/$c^2$ in the $M(\gamma J/\psi)$
distribution by assuming a flat distribution in the full mass
range. 

\begin{table}[htbp]
\begin{center}
\caption{The results on $e^+e^-\to \omega \chi_{c1,2}$. Listed in
the table are the product of radiative
correction factor, branching fraction and efficiency
$\mathcal{D}=(1+\delta^r)\cdot(\epsilon_{e}\cdot\mathcal{B}
(J/\psi\to e^+e^-)+\epsilon_{\mu}\cdot\mathcal{B}(J/\psi\to\mu^+\mu^-))$,
number of the observed events $N^{\rm obs}$,
number of backgrounds $N^{\rm bkg}$ in sideband regions,
and the upper limit (at the 90\% C.L.) on
the Born cross section $\sigma^{\rm B}_{\rm UL}$.} \label{Number}
\begin{tabular}{ccccccc}
\hline
         Mode    &$\sqrt{s}$ (GeV) & $\mathcal{D}~(\times10^{-2})$   & $N^{\rm obs}$ & $N^{\rm bkg}$  & $\sigma^{\rm B}_{\rm UL}$ (pb) \\
\hline
$\omega\chi_{c1}$  &4.31 &1.43    &  1    &$0.0^{+1.2}_{-0.0}$    & $<18$      \\
             &4.36       &1.27    &  1    &$1.0^{+2.3}_{-0.8}$      & $<0.9$     \\
             &4.39       &1.27    &  1    &$0.0^{+1.2}_{-0.0}$     &$<17$     \\
             &4.42       &1.25    &  0    &$0.0^{+1.2}_{-0.0}$     &$<11$    \\
\hline
$\omega\chi_{c2}$  &4.36 &0.95  &  5    &$1.0^{+2.3}_{-0.8}$    &$<11$     \\
             &4.39       &1.06   &  3    &$0.0^{+1.2}_{-0.0}$   &$<64$    \\
             &4.42       &0.98    &  2    &$0.0^{+1.2}_{-0.0}$     &$<61$   \\
\hline
\end{tabular}
\end{center}
\end{table}

The Born cross section is calculated from
\begin{equation}
\label{sigmab}
 \sigma^{\rm{B}}=\frac{N^{\rm {obs}}}
 {{\mathcal{L}}(1+\delta^r) (1+\delta^v)(\epsilon_1
 \mathcal{B}_1+\epsilon_2\mathcal{B}_2) \mathcal{B}_3},
\end{equation}
where $N^{\rm {obs}}$ is the number of observed signal events,
$\mathcal{L}$ is the integrated luminosity, $(1+\delta^r)$ is the
radiative correction factor which is obtained by using a QED
calculation~\cite{QED-Delt} and taking the cross section measured
in this analysis with two iterations as input, $(1+\delta^v)$ is the vacuum
polarization factor which is taken from a QED calculation~\cite{QED-Delt-V}.
For the $e^+e^-\to
\omega\chi_{c0}$~$[\omega\chi_{c1,2}]$ channel,
$\mathcal{B}_1=\mathcal{B}(\chi_{c0}\to\pi^+\pi^-)$ [$\mathcal{B}(J/\psi\to
e^+e^-)$],
$\mathcal{B}_2=\mathcal{B}(\chi_{c0}\to K^+K^-)$ [$\mathcal{B}(J/\psi\to \mu^+\mu^-)$],
$\mathcal{B}_3=\mathcal{B}(\omega\to \pi^+\pi^-\pi^0)\times
\mathcal{B}(\pi^0\to \gamma\gamma)$ [$\mathcal{B}(\chi_{c1,2}\to\gamma J/\psi)\times
\mathcal{B}(\omega\to\pi^+\pi^-\pi^0)\times
\mathcal{B}(\pi^0\to\gamma\gamma)$], and $\epsilon_1$ and $\epsilon_2$
are the efficiencies for the $\pi^+\pi^-$ [$e^+e^-$] and
$K^+K^-$ [$\mu^+\mu^-$] modes, respectively. 
For center of mass energies where the signal is not significant, we set upper limits at the 90\%
confidence level (C.L.) on the Born cross section~\cite{footnote}.
The Born cross section or its upper limit at each energy point for
$\EE\to \omega\chi_{c0}$ and $\EE\to \omega\chi_{c1,2}$ are listed
in Tables~\ref{CrossSection-up} and~\ref{Number},
respectively.

\begin{figure}[htbp]
\begin{center}
\includegraphics[width=0.45\textwidth]{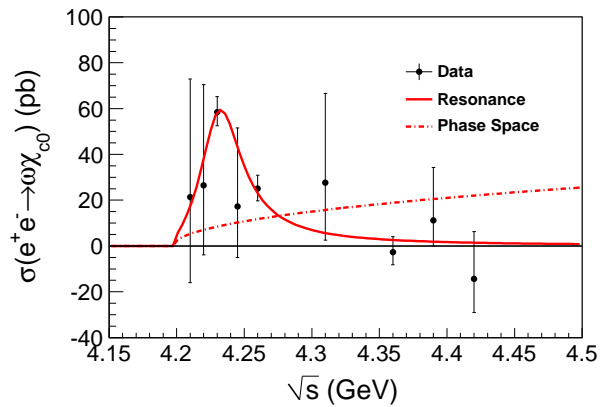}
\caption{Fit to $\sigma(e^+e^-\to\omega\chi_{c0})$ with a
resonance (solid curve), or a phase
space term (dot-dashed curve). Dots with error bars are the dressed
cross sections. The uncertainties are statistical only. }
\label{CS-BW-float}
\end{center}
\end{figure}

Figure~\ref{CS-BW-float} shows the measured Born cross sections
for $e^+e^-\to \omega\chi_{c0}$ over the energy region studied in
this work (we follow the
convention to fit the dressed cross section $\sigma^{\rm
B}\cdot(1+\delta^v)$ in extracting the resonant
parameters in~\cite{PDG}). A maximum likelihood method is used to
fit the shape of the cross section.

Assuming that the $\omega\chi_{c0}$ signals come from
a single resonance, a phase-space modified Breit-Wigner (BW) function
\begin{equation}
  {\rm BW}(\sqrt{s}) =
  \frac{\Gamma_{ee}\mathcal{B}(\omega\chi_{c0})\Gamma_t}
  {(s-M^2)^2+(M\Gamma_t)^2}\cdot\frac{\Phi(\sqrt{s})}{\Phi(M)}
\end{equation}
is used to parameterize the resonance, where $\Gamma_{ee}$ is the
$e^+e^-$ partial width, $\Gamma_t$ the total width, and
$\mathcal{B}(\omega\chi_{c0})$ the branching fraction of the resonance
decay to $\omega\chi_{c0}$.  $\Phi(\sqrt{s}) = \frac{P}{\sqrt{s}}$ is
the phase space factor for an $S$-wave two-body system, where $P$ is
the $\omega$ momentum in the $e^+e^-$ center-of-mass frame.  We fit
the data with a coherent sum of the BW function and a phase space term
and find that the phase space term does not contribute significantly.
The fit results for the resonance parameters are
$\Gamma_{ee}\mathcal{B}(\omega\chi_{c0})=(2.7\pm 0.5)$~eV,
$M=(4230\pm8)$~MeV/$c^2$, and $\Gamma_t=(38\pm12)$~MeV.
Fitting the data using the only phase space term results in a large
change of the likelihood [$\Delta(-2\ln L)=101.6$].  Taking the change
of 4 in the d.o.f.s into account, this corresponds to a statistical
significance of $> 9 \sigma$.

The systematic uncertainties in the Born cross section measurement
mainly originate from the radiative correction, the luminosity measurement, the detection
efficiency, and the kinematic fit. 
A 10\% uncertainty of in the radiative correction is estimated by varying
the line shape of the cross section in the generator from the
measured energy-dependent cross section to the $Y(4260)$ BW
shape.
Due to the limitation of the statistics, this item
imports the biggest uncertainty.
The polar angle $\theta$ of the $\omega$ in the $e^+e^-$
center-of-mass frame is defined as the angle between $\omega$ and
$e^-$ beam. For the $\omega\chi_{c0}$ channel, the distribution of
$\theta$ is obtained from data taken at 4.23~GeV and fitted with
$1+\alpha\cos^2\theta$. The value of $\alpha$ is determined to be
$-0.28\pm 0.31$. The efficiencies are determined from MC
simulations, and the uncertainty is estimated
by varying $\alpha$ within one standard deviation.
For the $\omega\chi_{c1,2}$ channels, a 1\% uncertainty is estimated by
varying the $\omega$ angular distribution from flat to $1\pm
\cos^2\theta$.
The uncertainty of luminosity is 1\%. The
uncertainty in tracking efficiency is 1\% per track.
The uncertainty in photon reconstruction is 1\% per
photon. A 1\% uncertainty in the kinematic fit is
estimated by correcting the helix parameters
of charged tracks~\cite{guoyp}.

For the $e^+e^-\to \omega\chi_{c0}$ mode, additional uncertainties come
from the cross feed between $K^+K^-$
and $\pi^+\pi^-$ modes, and the fitting procedure.
The uncertainty due to the cross feed is estimated to be 1\% by using
the signal MC samples.
A 4\% uncertainty from the fitting range
is obtained by varying the limits of the fitting range by $\pm
0.05~{\rm GeV}/c^2$.
The uncertainty from the mass resolution is
determined to be negligible compared to the
resolutions of the reconstructed $\omega$ in data and MC
samples. The uncertainties associated with $\mathcal{B}(\chi_{c0}\to
\pi^+\pi^-)$ and $\mathcal{B}(\chi_{c0}\to K^+K^-)$ are obtained to be 4\% by
varying the branching fractions around their world average values by
one standard deviation~\cite{PDG}.
A 5\% uncertainty
due to the choice of the background shape is estimated by changing the
background shape from the ARGUS function to a second order
polynomial (where the parameters of the polynomial are allowed to
float).
The overall systematic errors are obtained by
summing all the
sources of systematic uncertainties in quadrature
by assuming they are independent.
For the $\omega\chi_{c0}$ channel, they vary from 6.7\%
to 16.1\% depending on the center of mass energies.

The systematic uncertainties on the resonant parameters in the fit
to the energy-dependent cross section of $e^+e^-\to
\omega\chi_{c0}$ are mainly from the uncertainties of $\sqrt{s}$
determination, energy spread, parametrization of the BW function,
and the cross section measurement.
A precision of 2~MeV~\cite{beam} of the center-of-mass energy
introduce a $\pm 2$~MeV/$c^2$ uncertainty in the mass
measurement.
To estimate the uncertainty from the energy spread
of $\sqrt{s}$ (1.6~MeV), a BW function convoluted with a
Gaussian function with a resolution of 1.6~MeV is used to fit the
data, and the uncertainties are estimated by comparing the results
with the nominal ones. Instead of using a constant total width, we
assume a mass dependent width $\Gamma_t = \Gamma_t^0\cdot
\frac{\Phi(\sqrt{s})}{\Phi(M)}$, where $\Gamma_t^0$ is the width
of the resonance, to estimate the systematic uncertainty due to
signal parametrization. The systematic uncertainty of the Born
cross section (except that from $1+\delta^v$) contributes
uncertainty in $\Gamma_{ee}\mathcal{B}(\omega\chi_{c0})$. By
adding all these sources of systematic uncertainties in
quadrature, we obtain uncertainties of $\pm 6$~MeV/$c^2$, $\pm
2$~MeV, and $\pm 0.4$~eV for the mass, width, and the partial
width, respectively.

In summary, based on data samples collected between $\sqrt{s}=4.21$ and
4.42~GeV collected with the BESIII detector,
the process $e^+e^-\to \omega\chi_{c0}$ is observed at
$\sqrt{s}=4.23$ and 4.26~GeV for the first time, and the Born
cross sections are measured to be $(55.4\pm 6.0\pm 5.9)$ and
$(23.7\pm 5.3\pm 3.5)$~pb, respectively. For other energy points,
no significant signals are found and upper limits on the cross
section at the 90\% C.L. are determined.
The data reveals a sizeable $\omega\chi_{c0}$ production
around 4.23 GeV/$c^2$ as predicted in Ref.~\cite{omg-chi0}.
By assuming the $\omega\chi_{c0}$ signals come from a
single resonance, we extract the $\Gamma_{ee} \mathcal{B}
(\omega\chi_{c0})$, mass, and width of the resonance to be
$(2.7\pm 0.5\pm 0.4)$~eV, $(4230\pm 8\pm 6)$~MeV/$c^2$, and
$(38\pm 12\pm 2)$~MeV, respectively.
The parameters are inconsistent with those obtained by fitting
a single resonance to the
$\pi^+\pi^-J/\psi$ cross section
~\cite{babar-y4260-1}.
This suggests that the observed $\omega\chi_{c0}$
signals be unlikely to originate from the $Y(4260)$.
The $e^+e^-\to \omega\chi_{c1,2}$ channels
are also sought for, but no significant signals are observed;
upper limits at the 90\% C.L. on the production cross sections are determined.
The very small measured ratios of $e^+e^-\to\omega\chi_{c1,2}$ cross
sections to those for $e^+e^-\to\omega\chi_{c0}$ are inconsistent with the prediction
in Ref.~\cite{Ratio}.

The BESIII collaboration thanks the staff of BEPCII and the IHEP
computing center for their strong support. This work is supported
in part by National Key Basic Research Program of China under
Contract No. 2015CB856700; Joint Funds of the National Natural
Science Foundation of China under Contracts Nos. 11079008,
11179007, U1232201, U1332201; National Natural Science Foundation
of China (NSFC) under Contracts Nos. 10935007, 11121092, 11125525,
11235011, 11322544, 11335008; the Chinese Academy of
Sciences (CAS) Large-Scale Scientific Facility Program;
CAS under Contracts Nos. KJCX2-YW-N29, KJCX2-YW-N45; 100 Talents
Program of CAS; German Research Foundation DFG under Contract
No. Collaborative Research Center CRC-1044; Istituto Nazionale
di Fisica Nucleare, Italy; Ministry of Development of Turkey
under Contract No. DPT2006K-120470; Russian Foundation
for Basic Research under Contract No. 14-07-91152; U. S. Department
of Energy under Contracts Nos. DE-FG02-04ER41291, DE-FG02-05ER41374,
DE-FG02-94ER40823, DESC0010118; U.S. National Science Foundation;
University of Groningen (RuG) and the Helmholtzzentrum fuer
Schwerionenforschung GmbH (GSI), Darmstadt; WCU Program of National
Research Foundation of Korea under Contract No. R32-2008-000-10155-0.



\begin{thebibliography}{**}
\bibitem{footnote}The upper limit
is calculated by using a frequentist method with unbounded profile likelihood
treatment of systematic uncertainties, which is
implemented by a C++ class {\sc trolke} in the {\sc root}
framework~\cite{TRolke}. The number of the observed events is
assumed to follow a Poisson distribution, the number of background
events and the efficiency are assumed to follow Gaussian distributions. In
order to consider the systematic uncertainty in the calculation,
we use the denominator in Eq.~(\ref{sigmab}) as an effective
efficiency as implemented in {\sc trolke}.
\end{thebibliography}
\end{document}